\documentstyle[prl,twocolumn,aps]{revtex}
\input{epsf}
\begin{document}
\draft
\twocolumn[\hsize\textwidth\columnwidth\hsize\csname @twocolumnfalse\endcsname
\title{Dynamical turbulent flow on the Galton board with friction
and magnetic field}

\author{A. D. Chepelianskii$^{(a)}$ and D. L. Shepelyansky$^{(b)}$}

\address {$^{(a)}$ Lyc\'ee Pierre de Fermat, Parvis des Jacobins, 31068 
Toulouse Cedex 7, France}
\address {$^{(b)}$ Laboratoire de Physique Quantique, UMR 5626 du CNRS, 
Universit\'e Paul Sabatier, 31062 Toulouse Cedex 4, France}

\date{January 1, 2001; revised March 25, 2001}

\maketitle

\begin{abstract}
We study numerically and analytically the dynamics of charged particles on the 
Galton board, a regular lattice of disc scatters, in the presence of 
constant external force, magnetic field and friction.
It is shown that under certain conditions friction leads to the
appearance of a strange chaotic attractor. In this regime the average velocity 
and direction of particle flow can be effectively affected 
by electric and magnetic fields.
We discuss the applications of these results to the charge transport
in antidot superlattices and stream of suspended particles in a viscous
flow through scatters.
\end{abstract}
\vskip1pc]

\narrowtext

It is well known that dissipation can lead to the appearance of strange
chaotic attractors in nonlinear nonautonomous dynamical systems 
\cite{lieberman,ott}. In this case the energy dissipation is compensated 
by an external energy flow so that stationary chaotic oscillations 
set in on the attractor. Such an energy flow is absent in the Hamiltonian
conservative systems and therefore the introduction of dissipation 
or friction is expected to drive the system to simple fixed points in 
the phase space. This rather general expectation is surly true 
if the system phase space is bounded. However a much richer situation 
appears in the case of unbounded space, where unexpectedly a strange 
attractor can be induced by dissipation in an originally conservative 
system. 
To investigate this situation we study the dynamics 
of particles on the Galton board in the presence of constant external 
fields and friction. This board, introduced by Galton in 1889 \cite{galton},
represents a triangular lattice of rigid discs with which particles 
collide elastically.  For the case of free particle motion, the collisions
with discs make the dynamics completely chaotic on the energy surface 
as it was shown by Sinai \cite{sinai}. In this paper we study 
how the dynamics of a charged particle in a presence of
electric and magnetic fields is   
affected by a 
friction force ${\mathbf F_f} = - \gamma {\mathbf v}$ 
directed against particle velocity ${\mathbf v}$ (see Fig.~1). 
Without discs
an external in-plane electric field  $\mathbf{E}$ 
and a perpendicular magnetic field  $\mathbf{B}$ 
create a stationary particle flow with the velocity 
${\mathbf v_f} = ( {\mathbf f + F_L}) / \gamma$.
Here ${\mathbf{f}} = e {\mathbf{E}}$ is the effective force,
${\mathbf F_L} = e  \mathbf{ v_f \times } \mathbf{B } $ is the 
Lorentz force and
$e, m$ are the particle charge and mass.
All perturbations decay to this 
flow with a rate proportional to $\gamma$ 
so that this laminar flow can be considered as a
simple attractor.
The effects of friction inside one cell of the Galton board at 
${\mathbf B} = 0$ have been studied in $\cite{hoover}$ and it has been
found that friction leads to the appearance of a nontrivial
strange attractor. 
At present the effects of energy dissipation 
are actively investigated with the aim to
construct equilibrium and nonequilibrium steady states 
in a deterministic way (see \cite{hoover1,klages} and Refs. therein).
Here the Nos\'e-Hoover thermostat 
with a momentum-dependent friction coefficient
allows to reach a number of interesting results
with applications to molecular dynamics and nonequilibrium liquids
\cite{hoover1}. In our studies, contrary to \cite{hoover},
we concentrate mainly on the spatial structure of 
the turbulent chaotic particle flow appearing in the presence of
friction. We show that the flow direction can be efficiently affected by 
a magnetic field. The obtained results describe   the electron 
dynamics in antidot superlattice which has been experimentally realized
in semiconductor heterostructures \cite{weiss}. In such structures the 
effects of classical chaos play an important role \cite{geisel} and the
effects of friction we discuss here can appear for relatively strong
electric fields. 
\vglue 0.0cm
\begin{figure}
\epsfxsize=8cm
\epsfysize=7cm
\epsffile{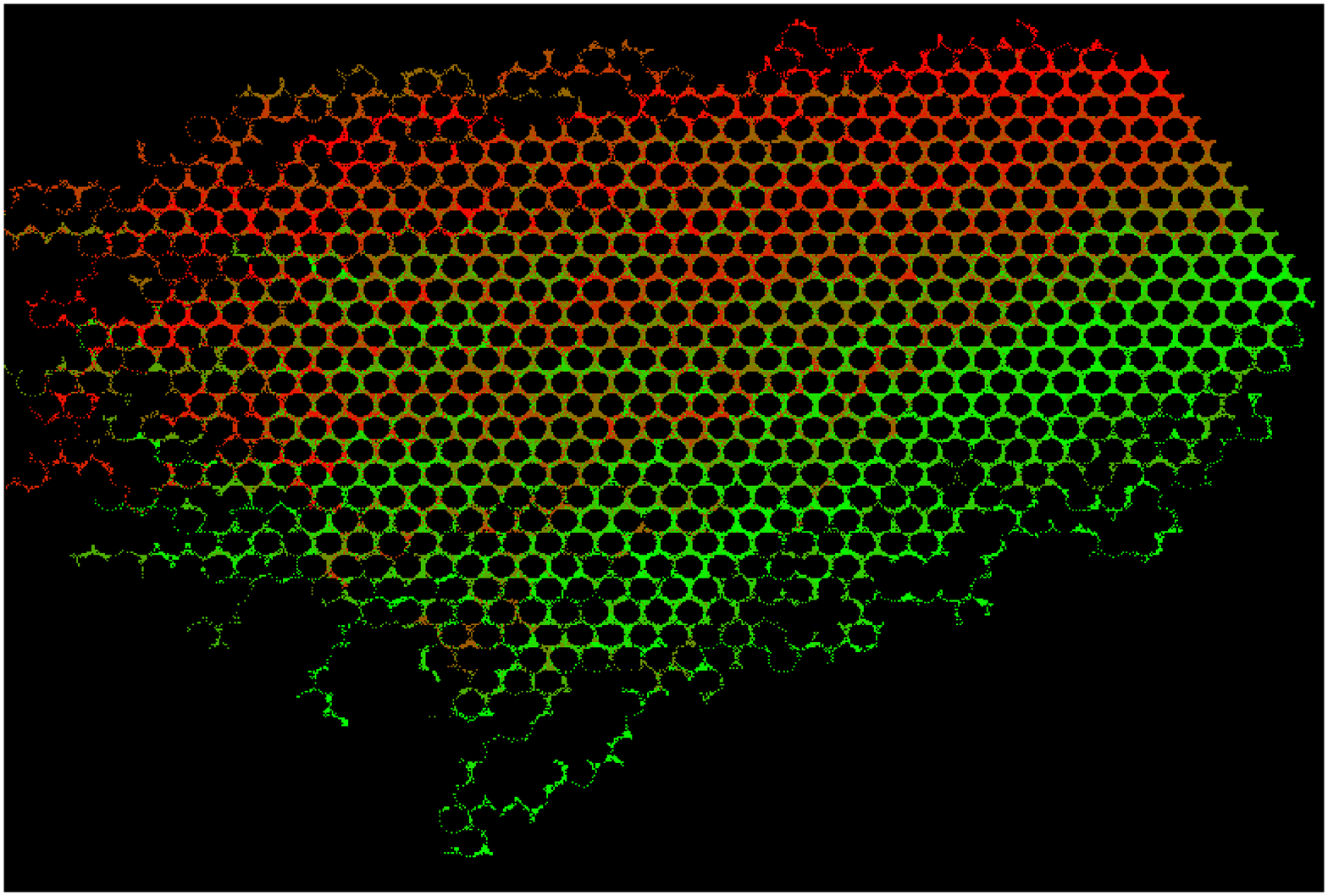}
\vglue 0.2cm
\caption{Chaotic flow on the Galton board. Here the distance 
between discs is $R = 2.24$, ${\mathbf f}$ = (-0.5, -0.5),
$B = 2$ and $\gamma = 0.1$.
Initially 200 particles are distributed homogeneously along 
a straight line segment in the upper right corner, their
color homogeneously changes from red to green along this segment.
} 
\label{fig1}
\end{figure}

To study numerically the dynamics of this model, we fix 
the disc radius $a = 1$,
and  $e = m = 1$ so that the system is 
characterized only by the distance $R$ between 
discs packed into equilateral triangles all over the $(x,y)$ plane,
the friction coefficient $\gamma$, the external force of strength 
$f=|{\mathbf f}|$ and the cyclotron frequency $\omega_c = B$
\cite{note}. 
For $R \le R_c = 4/\sqrt{3}$ there is no straight line, by 
which a particle can cross the whole plane without collisions
(at $f = 0, B=0, \gamma = 0$), and we start the discussion from this case.
The particle dynamics is simulated numerically by using the exact 
solution of Newton equations between collisions, and by determining 
the collision points with the Newton algorithm. This way the 
trajectories are computed with high precision, and for example 
for $\gamma = 0$ the total energy is conserved with a relative 
precision of $10^{-14}$ \cite{note1}. A typical example of the chaotic flow
formed by an ensemble of particle trajectories is shown in Fig.~1.
To illustrate the mixing properties of the flow we attributed
a color to each trajectory that allows to follow their interpenetration
and spreading. This figure shows that there is a certain penetration
depth of one color into another, however this depth is finite since 
on average the initial color repartition is still visible.
\begin{figure}
\epsfxsize=7.0cm
\epsfysize=6.0cm
\epsffile{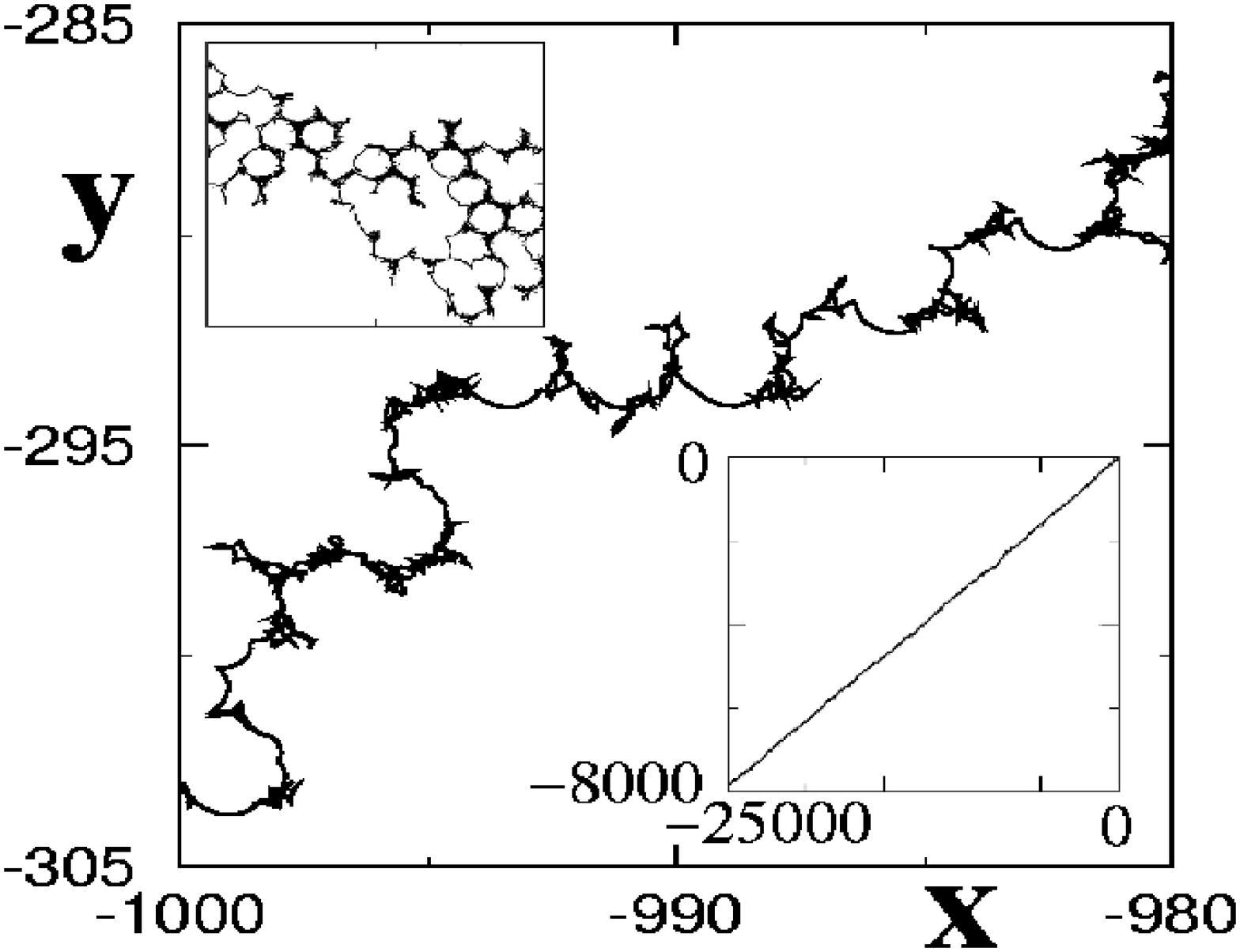}
\vglue 0.2cm
\caption{ Example of a single trajectory for the case of Fig.~1
shown on small (main figure) and large (lower inset) scales.
The drift velocity of the flow $v_f \approx 0.13$
is directed at angle $\alpha \approx 0.48$ to ${\mathbf f}$.
Upper inset shows a single trajectory in 
the region $(-160 \leq x \leq -140, -190 \leq y \leq -170)$ for 
$\gamma=0.004$ with $v_f \approx 0.05$.
} 
\label{fig2}
\end{figure}

The properties of color penetration can be understood from the 
analysis of single trajectory dynamics. Such a typical example
is presented in Fig.~2. It shows that the particle moves with
an average constant velocity $v_f$ under some angle $\alpha$ to the
external force ${\mathbf f}$, except for $B=0$ where $\alpha =0$.
This drift velocity is constant only on average
since on a smaller scale the particle moves chaotically between
scatters following a strange chaotic attractor. 
In Fig.~2 for $\gamma=0.1$ the drift velocity is relatively 
large and the particle does not have enough time to move around 
many scatters in the direction perpendicular to the flow. 
As a result the penetration depth for color mixing is not very large. 
Surprisingly, for smaller friction, the drift velocity becomes smaller
and the penetration depth increases so that the particle makes many 
turns around discs as it is shown in Fig.~2 for $\gamma=0.004$.
This dependence is opposite to the case without
scatters where $v_f$ drops with the $\gamma$ growth.

This result can be understood on the basis of the following
physical arguments (for $B=0$ see also \cite{hoover}).
In the regime with weak friction the particles start to diffuse 
among discs in a chaotic manner with the diffusion rate 
$D = v l/2$ where $v$ is the particle velocity  and $l$ 
is the mean free path \cite{ziman}.
For $R \sim 1$ we have  $l \sim R \sim 1$
while the dependence of $l$ on $R$ will be discussed in more detail
latter. During the dissipative time scale $\tau_{\gamma} = m / \gamma$
this diffusion leads to the particle displacement 
$\Delta r \sim \sqrt{D m/ \gamma}$
along the direction of the drift velocity ${\mathbf v_f}$. 
This gives the change in the 
potential energy 
$U \sim f \Delta r \cos \alpha  \sim f_{ef} \sqrt{D m/ \gamma}$
where $f_{ef} = f \cos \alpha$.
In the stationary regime at time $t \gg \tau_{\gamma}$, this potential
energy should be comparable with the kinetic energy of the particle so
that $U \sim m v^2$, where $v^2 =  <v^2_x + v^2_y>$
is the average velocity square.
Hence
\begin{equation}
\label{velav}
v^2 \sim ( f \cos \alpha)^{4/3} ( l / \gamma m)^{2/3} \;\; .
\end{equation}
This relation  allows to determine 
the drift velocity of the flow 
${\mathbf v_f}$. Indeed during the time $\tau_c$ 
between collisions the particle is accelerated by average forces ${\mathbf f}$
and ${\mathbf F_L} = e {\mathbf v_f \times B}$ that gives 
the average drift velocity ${\mathbf v_f} = ({\mathbf f + F_L}) \tau_c / m$.
Since the dynamics is chaotic the direction of 
velocity is changed randomly after each collision 
so that ${\mathbf v_f}$ is accumulated only between collisions. 
The time $\tau_c$ is determined by the mean 
free path $l$ and the average velocity $v$:
$\tau_c \; = l / v \; = 2 D / v^2$. 
Thus for the angle $\alpha$ between ${\mathbf v_f}$ and ${\mathbf f}$
we obtain the relation:
\begin{equation}
\label{tan}
\tan \alpha = e B \tau_c/m \sim { {e B \; l^{2/3} \; \gamma^{1/3}} 
\over {(m f \cos \alpha )^{2/3} }} \; .
\end{equation}
The amplitude of fluctuations around this direction is
$\Delta r \sim \sqrt{D m / \gamma}$ 
which also determines the color mixing depth (see Fig.~1).

From (\ref{velav}) and (\ref{tan}) we obtain the drift velocity amplitude
\begin{equation}
\label{veldrift}
v_f = f_{ef} \tau_c/m \sim l^{2/3} (\gamma f_{ef} /m^2)^{1/3} \;\;,
\end{equation}
with $f_{ef} = f \cos \alpha$.
For $B=0$ the particles flow in the ${\mathbf f}$ direction.
In this case their mobility is $\mu = v_f / f = \tau_c/m = D / (m v^2/2)$. 
This is in fact the Einstein relation according to which the mobility 
is given by ratio of the diffusion rate to the average kinetic 
energy (temperature) \cite{landau}. 
At $B=0$ the relations (\ref{velav}), (\ref{tan}) and (\ref{veldrift})
are in the agreement with those in \cite{hoover}
and with the numerical data shown in Fig.~3.
The values of $v_f$ and $v^2$ are obtained from one very long trajectory
(with a length of $10^3 - 10^4 \; R$ )
or 10 shorter trajectories. Within statistical fluctuations
this gives the same  $v_f$ and $v^2$ independent of their initial values.

The relations (\ref{velav}), (\ref{tan}), (\ref{veldrift}) allow to estimate 
the value of the Lyapunov exponent $\lambda$. Indeed, the particle 
moves with a typical velocity $v$ and as in the case of the Sinai 
billiard with $l \sim R \sim 1$ we have $\lambda \sim v / l$.
Therefore  in the regime when:
\begin{equation}
\label{gammac}
\gamma < \gamma_c \sim \sqrt{m f_{ef}  / l} \;\; ,
\end{equation}
the value of $\lambda$ is much larger than the dissipation rate $\gamma / m$.
As a result for $\gamma \ll \gamma_c$ the strange attractor is fat 
and its fractal dimension is close to the maximal dimension $4$,
which is determined by the number of degrees of freedom (we remind 
that contrary to the nondissipative case the energy is not conserved).
For $\gamma \gg \gamma_c$ the dissipation time $\tau_{\gamma}$ becomes 
much shorter than the time between collisions $\tau_c$. In this case 
the dissipation dominates chaos and the strange attractor degenerates into
a simple attractor. For $\gamma < \gamma_c$ our numerical simulations
performed with high computer accuracy show that  trajectories remain
chaotic for displacements from the origin
being larger than $10^5 R$. Also at $B=0$
it can be shown that  
$\gamma_c \sim m v/l \sim \sqrt{ f m a } / R$ for $R \gg a$
and $\gamma_c  \sim \sqrt{m f/ a^3} \; \Delta R$ for
$\Delta R = R - 2a \ll a$.

\begin{figure}
\epsfxsize=8cm
\epsfysize=6cm
\epsffile{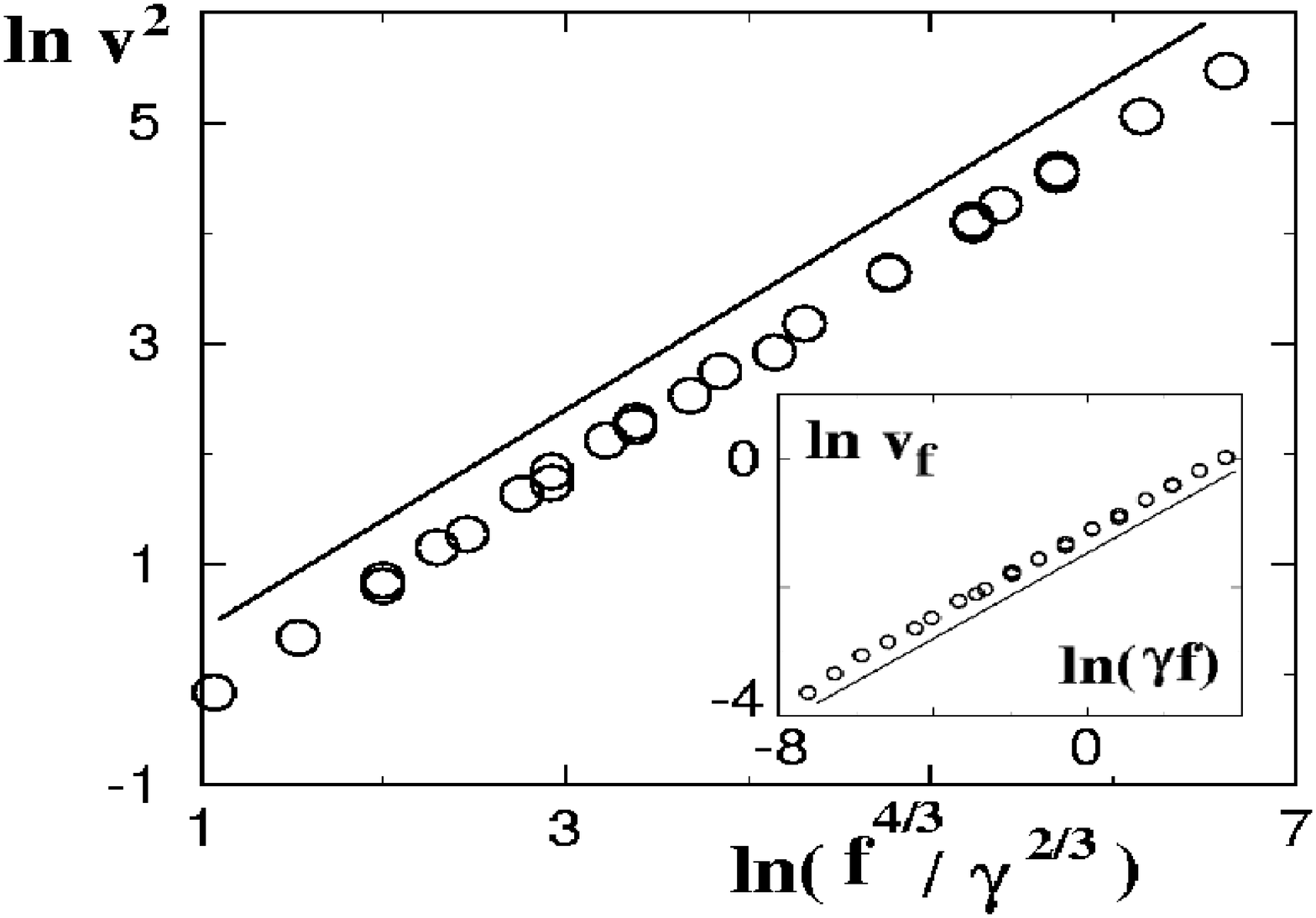}
\vglue 0.2cm
\caption{Dependences of
$v^2$  and $v_f$ (inset) on $f$ and $\gamma$
for $R=2.24$, ${\mathbf f}/|{\mathbf f}| = (-1,-1)$, $B=0$
and $0.001 \leq \gamma \leq 0.4$; $0.5 \leq f \leq 32$:
circles show numerical data and lines show the 
slopes from (\ref{velav}) (main figure) and (\ref{veldrift}) (inset).
} 
\label{fig3}
\end{figure}

The above changes in the mean free path $l$ at $B=0$ also affect the drift 
velocity of the flow through the relation (\ref{veldrift}).
Indeed, for $\Delta R \ll 1$ we expect $l \sim \Delta R$ that 
gives $v_f \propto {\Delta R}^{2/3}$. This dependence is close 
to the numerical data shown in Fig.~4  even if the numerical 
value of the exponent is approximated better by $0.5$.
In the other limit $R \gg 1$, we have $l \sim R^2 / a$ that gives 
$v_f \propto R^{4/3}$. This power dependence is in a satisfactory 
agreement with the data in Fig.~4 although 
the numerical value of the exponent is closer to $1$.
We attribute these small deviations in the exponent values to 
a restricted interval of variation in $R$. Actually we can not 
use very large/small values of $\Delta R$ since in these limits 
the value of $\gamma$ becomes comparable with $\gamma_c$
and chaotic attractor disappears. We note that according to our 
data (see Fig.~4) the strange attractor exists even in the case 
$R > R_c = 4/\sqrt{3}$ when at $f = 0, \gamma = 0$ there are straight
trajectories crossing the whole plane without collision.
Apparently the contribution of these orbits is not significant 
if $\gamma > 0$ and if $\mathbf{f}$ is not directed along these lines.

\begin{figure}
\epsfxsize=7.5cm
\epsfysize=6.5cm
\epsffile{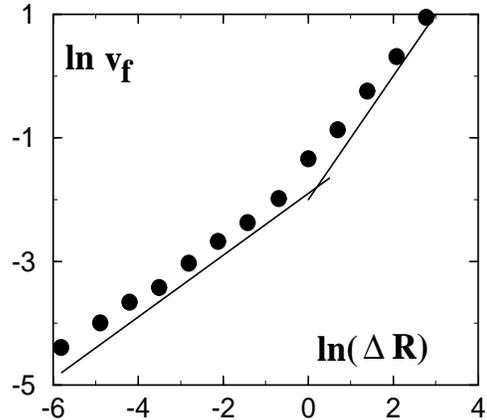}
\vglue 0.2cm
\caption{Dependence of $v_f$ on $\Delta R =R -2$ for $\gamma=0.1, B=0$
and ${\mathbf f} =(-0.5,-0.5)$: points give numerical data, lines show
the slopes 0.5 and 1. 
} 
\label{fig4}
\end{figure}

\vskip -0.5cm
\begin{figure}
\epsfxsize=8.7cm
\epsfysize=6.5cm
\epsffile{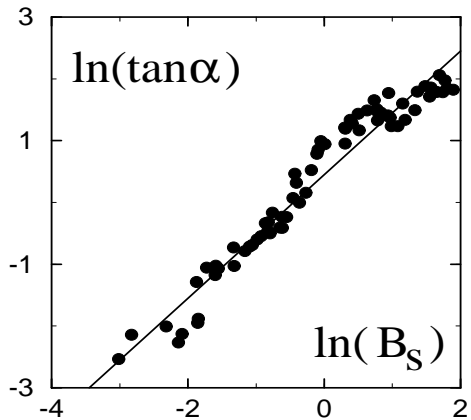}
\vglue 0.2cm
\caption{Dependence of $\tan \alpha$ on scaled magnetic
field $B_s=B \gamma^{1/3} {f_{ef}^{-2/3}}$ from (\ref{tan})
with $f_{ef}=f \cos \alpha$ for $R=4$, ${\mathbf f}/|{\mathbf f}| =
(-1,-1)$ and
$0.125 \leq B \leq 16$, $0.14 \leq  f \leq 5.65$,
$ 0.002 \leq \gamma \leq 0.1$. Points give numerical data,
line shows the average dependence $\tan \alpha = 1.6 B_s$.
} 
\label{fig5}
\end{figure}

The introduction of magnetic field allows to change efficiently
the direction of the flow. The numerical data for the 
variation of $\tan \alpha$ with the magnetic field and 
other system parameters are presented in Fig.~5.
The average dependence is in a good agreement with the equation (\ref{tan})
for a large region of parameter variation where $\tan \alpha$ changes 
by two orders of magnitude.
At the same time for moderate angles $\alpha < 1$ the flow 
velocity $v_f$ is weakly affected by $B$. For example, for the case of Fig.~2,
$v_f$ remains practically the same for $B = 2$ and 
$B = 0$ ($v_f \approx 0.15$). For $\alpha > 1$ magnetic field 
starts to change significantly $v^2$ and $v_f$ in agreement with 
(\ref{velav}) and (\ref{veldrift}).
While the data in Fig.~5 in average follow the dependence (\ref{tan})
the fluctuations around the average are rather large.
Their origin becomes clear from Fig.~6 where only the magnetic field 
is changed. In this case, $\tan \alpha$ has a pronounced peak
which is located near the value of $B$ where the cyclotron radius 
$r_c = v / B$ is equal to the disc radius. Indeed for $r_c > a$ 
a trajectory can make a full turn around a disc that allows to increase 
$\alpha$ and to reach a strong deviation of the global flow from 
the direction of the electric field. The growth of $\alpha$ leads 
to a drop of currant (conductivity) in the ${\mathbf f}$ direction
and hence to the increase of resistivity. In fact, the peaks in 
resistivity near $r_c \approx a$ had been observed experimentally
\cite{weiss} and explained theoretically \cite{geisel} in the 
linear response regime. Our data show that the peaks should 
also exist in the regime with strange chaotic attractor and 
relatively strong electric field
where the $I-V$ characteristics become nonlinear.
For a denser package of discs the above peak is still present 
($R = 2.24$ in Fig.~6) but it is less pronounced.
It is interesting to note that in this case, $\alpha$ can even be 
negative so that the particles flow against the average Lorentz force.
We note that the possibility of such a flow had been discussed for systems 
with Hamiltonian dynamical chaos in \cite{geisel1}. 
Generally for a dense disc package the contribution
of resonant orbits with $r_c \approx a$
starts to be significant and deviations from the 
average dependence (\ref{tan}) become rather strong.

\vskip -0.3cm
\begin{figure}
\epsfxsize=8.7cm
\epsfysize=6.5cm
\epsffile{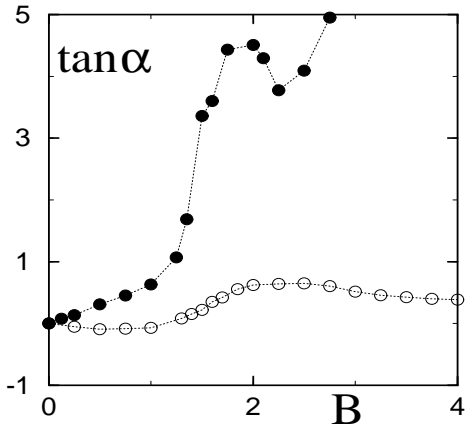}
\vglue 0.2cm
\caption{Dependence of $\tan \alpha$ on $B$
for ${\mathbf f} =(-0.5, -0.5)$, $\gamma =0.03$,
$R=4$ (points) and $R=2.24$ (circles). Cyclotron and disc 
radiuses are equal at $B \approx 1.6$ where $v^2 \approx 2.5$.
} 
\label{fig6}
\end{figure}

The above dynamical turbulent flow has rather interesting and unusual 
properties and it would be interesting to study it in experiments with 
antidot superlattices like in \cite{weiss}. The regime we discussed 
should appear when the steady state velocity 
$v$ from (\ref{velav}) becomes larger than the Fermi velocity $v_F$
of the charge carriers. Thus, in addition to
(\ref{gammac}), the condition $v > v_F$ determines the 
threshold for appearance of a strange attractor.
The experimental investigation of this phenomenon should also 
shade light on the role of quantum effects in such regime.
We note that in the derivation of (\ref{velav}) - (\ref{gammac})
we didn't use any specific properties of the disc distribution in the 
plane and therefore the results remain valid for randomly
distributed discs. 
Our study represents also certain interest for
transport properties of neutral/charged particles suspended in a viscous flow 
streaming through a system of scatters. Indeed,  
a laminar stream with the velocity $v_{s}$ creates an effective 
force $f = v_{s} {\gamma}_{ef}$ where ${\gamma}_{ef}$ is the effective
friction created by the viscosity of the liquid.
Such kind of transport can be studied experimentally with viscous
liquids and its investigation can contribute to a better understanding 
of the interplay between dissipation, turbulence and chaos.

We thank R. Klages for constructive critical remarks and 
for pointing to us Refs. \cite{hoover,hoover1,klages}, and
W. Hoover for his stimulating interest to our work.


\begin{thebibliography}{99}
\bibitem{note} Dimensional analysis shows that
        one of these 4 parameters can be fixed
        but to trace their physical origin
        we keep  all of them.
\bibitem{note1} Between collisions the energy is not alternated
        while Newton method converges rapidly to thereof precision.
\end{thebibliography}
\end{document}